# Exploring the legacy of big stargazing events

*This article was selected as a Free Editors Choice Article in the December 2018 issue of A&G, and is available to download fully typesetted at:* [https://doi.org/10.1093/astrogeo/aty278](https://doi.org/10.1093/astrogeo/aty278)

**Karen L Masters**, **Jennifer A Gupta** and **Wiktoria Kedziora** assess the impact of annual large- scale stargazing events, following the success of *Stargazing Live* and based on their experiences in Portsmouth.

The launch of the BBC's award- winning *Stargazing Live* TV programmes in January 2011 kick-started a revolution in the quantity and focus of annual public stargazing events right across the UK. While observatories, universities, amateur astronomical societies and others had been organizing public astronomy/observing events for many years, the programme provided both a focal point for these events and an opportunity to bring in new audiences.

From 2011 to 2015, the BBC actively encouraged organizations to put on events that tied in with *Stargazing Live*, providing free supplementary educational and advertising materials, suggestions for activities, and how-to guides for running large-scale events. As a result, hundreds of stargazing events took place all over the UK around the time of the TV broadcast, all listed on the BBC "Things To Do" website. BBC Learning also directly organized regional, and then national, partner events in the early years, from which small segments appeared on the national TV show. This created local stargazing events, aimed at beginners, on a scale never seen before. It is reported, for example, that "40 000 people attended 330 linked events nationwide" in 2011 (cited by the University of Manchester 2014 Research Excellence Framework [REF] submission), while 113 000 people attended events in 2012 (according to the BBC Media Centre website). These events had impact; the Manchester REF2014 impact case study noted, for example, that "astronomy equipment was reported as selling out in major retailers" and "astronomy societies reported significant boosts in membership and website traffic".

The TV programme won the Broadcast Digital Award for Best Content Partnership in 2014 and was shortlisted again in 2016. But its future is unknown: there was no broadcast from the UK in 2018 and no announcements have been made about 2019 at the time of writing. Despite this, many astronomy organizations (amateur, museum and university based) now consider an annual stargazing event a staple component of their outreach and public engagement programmes. The Institute of Cosmology and Gravitation (ICG) at the University of Portsmouth is one of them. It is our experience that the success of the earlier tie-in events has left a lasting legacy in our local community, and members of the public are now conditioned to expect and look forward to public stargazing events in January. Here we look at the longer-term impact of the local tie-in events, using the events we have organized in Portsmouth as a case study.

**Portsmouth events**
The first Portsmouth-based *Stargazing Live* event, organized by members of the ICG, was relatively modest in scale. In 2011, we took a few telescopes to the local shopping centre and distributed materials provided by the BBC. The following year, BBC South used the Spinnaker Tower in Portsmouth as a location for its BBC Learning event for the South region, including small segments shown on the main TV show (and regional news broadcasts from the site). Members of the ICG were the main academic partners for this event.

The local success of this event encouraged us to run an annual large-scale – and free – *Stargazing Live* tie-in event in Portsmouth, and to work with our local amateur astronomy society (Hampshire Astronomical Group) and other organizations to do this. A shift was made to hosting the event at attractions within Portsmouth Historic Dockyard, specifically HMS Warrior 1860 (2013 onwards), the now-defunct Mary Rose Story (2013) and Action Stations (2014 onwards), organized in collaboration with the hosts' education teams. Despite poor weather in 2013, 450 members of the public attended. The following year, BBC Learning selected Portsmouth Historic Dockyard as the location for one of its three national events, which made it a much larger event – an order of magnitude greater than the capacity of locally run events. The University of Portsmouth contributed to this event as the lead academic partner, but many other groups of professional astronomers were also involved (e.g. the Universities of Southampton and Oxford).

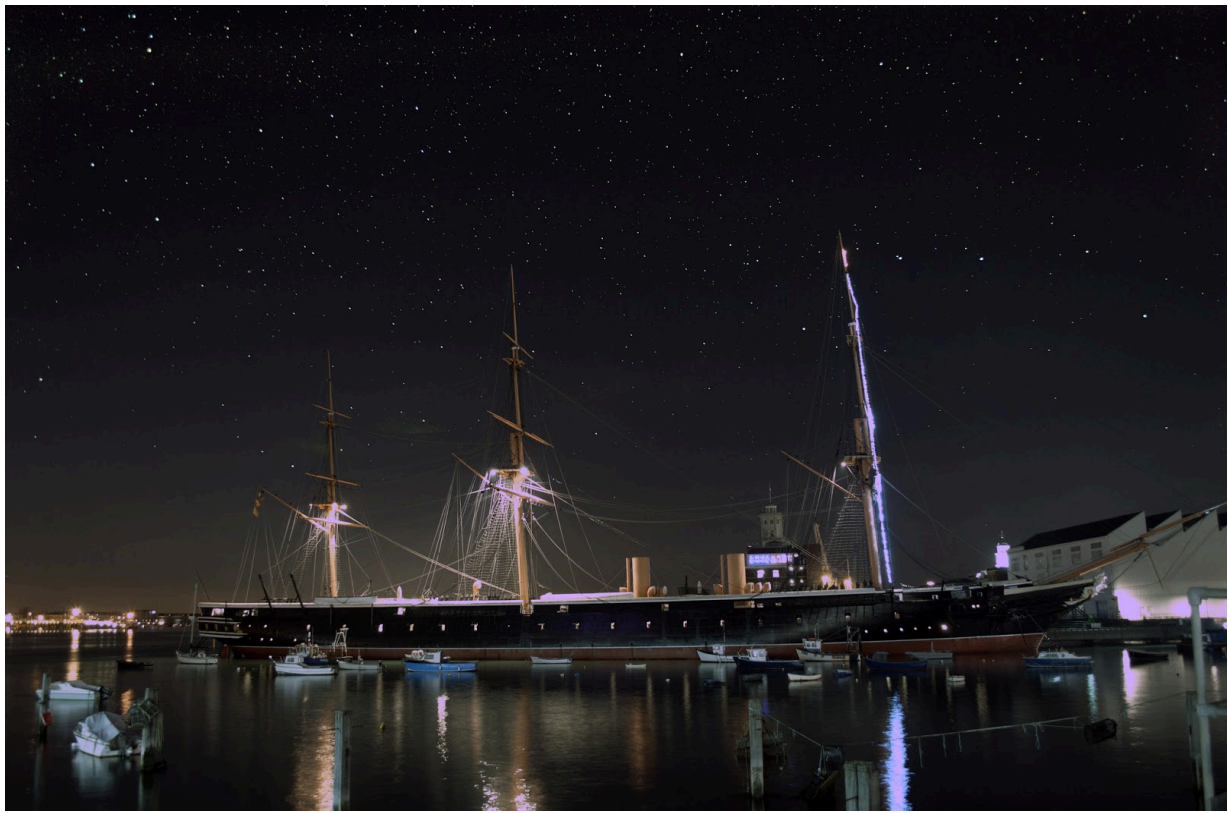

Stars over HMS Warror 1860. This composite image, taken with an 18 mm lens, uses a stack of ninety 20 s images (aligned to remove star trails) to reveal the background stars that appear behind a single short image of the foreground taken from the same location. (Edd Edmondson)

Following this BBC-led event, we have continued to run our own Stargazing at Portsmouth Historic Dockyard events, with evening sessions held in March 2015 (plus an additional daytime event in Portsmouth Guildhall Square for the partial solar eclipse) and in January 2016 to coincide with *Stargazing Live* in those years. With no further active support from the BBC, and no evidence that attendance at our local events were tied to the timing of the TV show, we decided to run the event again in January

2017 and 2018. *Stargazing Live* was broadcast in March 2017 and in 2018 was not hosted in the UK. We plan an event in Portsmouth on 30 January 2019, although at the time of writing we do not know if the BBC *Stargazing Live* will happen in 2019.

Attendance at these events has grown steadily from about 450 in 2013 to more than 800 in 2018; the BBC-led events in 2012 and 2014 are in a different league, with around 5500 members of the public attending in 2014.

Our Stargazing at Portsmouth Historic Dockyard events include a lot more than stargazing. The collaboration with HMS Warrior 1860 and Action Stations (now both managed by the National Museum of the Royal Navy, NMRN), along with the ICG's research strengths in cosmology, led to events with themes around mapping and navigation – of the world and the universe. The most recent event in January 2018 included a range of activities for all ages and abilities, including a series of talks by members of the ICG, hands-on activities to explain ICG research such as gravity and galaxy evolution, stands where attendees could ask ICG researchers questions, astronomy-themed arts and crafts and face-painting, rocket building with Action Stations, astronavigation activities with NMRN, a telescope clinic with Hampshire Astronomical Group (the weather meant that no actual stargazing took place) and space engineering activities with Airbus Defence and Space. The emphasis on providing indoor activities is by no means unique to us, but is instrumental in the ongoing success of the events regardless of the weather.

**Long-term impact?**

There are many reasons why a university department may want to organize an event like Stargazing at Portsmouth Historic Dockyard. They allow for engagement with the local community, breaking down barriers between the university and those who live nearby. They provide an opportunity for researchers to gain experience in public engagement and can also be an excellent team-building experience. However, it is important to consider who the intended audience is for such events, and the impact that the event has on them.

In a public engagement context, the term "impact" is normally used to describe any longer-term change that has come about as a result of the engagement. This means looking at how the attendees have been affected in the long-term: for example, have they remembered information from the event (change in knowledge), or has it inspired them to seek out similar events or activities in the future (change in behaviour)? It is important to monitor and analyse this impact because it gives an indication of whether the event is serving its purpose. It also provides objective evidence when seeking support for future events. Furthermore, the growing importance of "Impact" within the REF assessment has made such evidence even more valuable.

In every year (excluding 2012 and 2014 where ticketing was managed by the BBC), we collected immediate feedback from attendees through a questionnaire that was available on the night and emailed to attendees shortly after. Each year this feedback was extremely positive, revealing that the overwhelming majority of attendees had enjoyed the event, and eliciting quotes from attendees such as: "[A] very well organized event. My daughter and I thoroughly enjoyed it. Normally, she is not that interested in science but she loved all of the activities." Furthermore, when asked the question "Having attended this event, are you now more likely to do any of the following?", more than 80% responded positively to "learn more about astronomy" and "visit a science museum, observatory or planetarium" and 95% responded positively to "attend future public events organized by the ICG" fol- lowing the 2016 event. These responses indicate that the events should have had a longer-term impact on attendees, changing their behaviour and attitude towards astronomy. However, we wanted to collect more robust evidence that would look at the long-term impact of running an event in the same, or similar, location every year for several years. This led us to apply for, and obtain, Small Award funding from STFC to run a small longitudinal study of people who had attended our events over the past several years.

As a result, in the summer of 2017 a follow-up questionnaire was sent out to the attendees of events held in 2013, 2015, 2016 and 2017; 2012 and 2014 were run by the BBC so we had no contact details for attendees, and no records were kept in 2011. We had 123 responses, with representation from all events within the time period 2012–17, despite no 2012 or 2014 attendees being contacted directly. This already indicates some repeat attendance. The survey distribution and analysis were led by Wiktoria Kedziora, a University of Portsmouth physics undergraduate student, as part of a summer placement project supervised by Karen Masters and Jen Gupta. A small prize draw, for a signed book by Chris Lintott, was used to encourage a response.

Overall, the responses to the questionnaire demonstrate that our events have had a positive impact on the attendees. Almost 70% of respondents answered that Portsmouth Stargazing events had increased their interest in astronomy, with a similar number reporting they had gone on to attend similar events, or participate in other astronomy-related activities. Furthermore, more than half of respondents had been to other science-related events following attendance at a Portsmouth Stargazing event, with 45% of these attendees reporting an increase in attending such events. Three-quarters of the respondents also reported that they intend to attend a future Portsmouth Stargazing event.

Some respondents did report that the Stargazing events enabled them to find out about other astronomy-related organizations etc; for example, one attendee reported: "[I] wasn't aware of the [Hampshire Astronomical Group] Observatory in Clanfield until the last event at Portsmouth Dockyard." But in the past, we made only limited deliberate efforts to direct attendees to specific follow-up events or activities. This is reflected in the fact that several individuals reported being unsure of how to further their interest in astronomy beyond attending our events. There is a clear need to provide such information, in order to give our annual event a higher and longer-term impact.

A handful of attendees reported that they or their children had gone on to further study in astronomy (or physics) after the event they attended, although this cannot be claimed as a direct result. One notable demographic gap in attendance currently limits this potential impact – while young children (up to early secondary school age) and over-30s were well represented, there were very few teenagers or young adults who might be going on to university soon. How- ever, the ICG runs a separate outreach programme to target school and college pupils, which is likely to be more effective than a one-off event in terms of encouraging more young people to study physics and astronomy.

As well as looking at the impact on attendees in terms of astronomy and stargazing, we were also keen to look at the effect of the location of the events. Portsmouth Historic Dockyard is an unusual venue for a stargazing event, but 90% of respondents reported that they had previously visited it prior to a Portsmouth Stargazing event, and 57% reported that they had visited since. This suggests that the Stargazing at Portsmouth Historic Dockyard events are attracting people who would visit the dockyard anyway. While many respondents provided positive comments about the venue (particularly HMS Warrior 1860), the overwhelming majority (81%) said that holding the event at a different venue would make no difference in their likelihood of attending in the future.

Analysis of the demographic information of attendees who responded to the questionnaire reveals that only around half (48%) had no connection to the University of Portsmouth; the remainder comprised current and former staff and students, or their family members. Postcode data were also gathered, with all postcode districts within Portsmouth, Havant, Gosport and Fareham represented, although in some cases there was only one response from a district. This was used for a basic analysis of socioeconomic status, using information from the website http://www.postcodearea.co.uk. More than a quarter of respondents were from the PO4 and PO5 postcode areas, which are dominated by middle class households (69% and 62% respectively). In contrast, only 6.5% of respondents were from the PO1 postcode area, which is dominated by "working class" and "non-working" households (a total of 64%) indicating lower socioeconomic status, despite the event taking place in this part of Portsmouth. While this is perhaps not surprising, it has led us to consider how best to work with underserved communities in Portsmouth, something we are keen to develop over the 2018/19 academic year.

**Conclusion**

Large-scale stargazing events have become part of the landscape of astronomy public engagement in the UK, following a period of four years where events linked to the BBC's *Stargazing Live* TV programme were actively encouraged. We used the annual events run in Portsmouth by the University of Portsmouth's Institute of Cosmology and Gravitation as a case study, to look at the longer-term effect of these events on the attendees. We found consistent self-reporting of increased interest in astronomy following our events, as well as changes in behaviour. But one-off events such as these need to provide clear guidance about follow-up activities if they are to have a longer-term impact. We also find evidence that these types of large-scale events are not the best way to appeal to lower socio- economic status households locally.

The impact initiated by *Stargazing Live* and tie-in events led by astronomers across the UK has been significant. Without this trigger point it seems unlikely that the University of Portsmouth would have committed to such large-scale annual stargazing events, and by incorporating activities that directly engage attendees with the research done at the Institute of Cosmology and Gravitation, the events have been a great opportunity for us to engage with our own local community about the research done in our department.

**AUTHORS**
Karen L Masters is Associate Professor of Physics and Astronomy at Haverford College, Pennsylvania, USA; and Reader at University of Portsmouth, UK. Jennifer A Gupta is Outreach Manager at University of Portsmouth. Wiktoria Kedziora is a physics undergraduate at University of Portsmouth.

**ACKNOWLEDGMENTS**
An STFC Small Award in 2016: https://stfc.ukri.org/public-engage- ment/public-engagement-grants/pe-funding-opportunities/pe-small-awards/2016b-successful-projects